\documentstyle[twocolumn,aps]{revtex}
\begin{document}

\title{Induced vs Spontaneous Breakdown of S-matrix Unitarity:
\\ Probability of No Return
in Quantum Chaotic and Disordered Systems}
\author{Yan V. Fyodorov}
\address{Department of Mathematical Sciences, Brunel 
University, Uxbridge UB83PH, United Kingdom}
\address{Petersburg Nuclear Physics Institute RAS,
 188350 Gatchina, Leningrad Reg., Russia}

\date{\today}
\maketitle

\begin{abstract}
We investigate systematically sample-to sample fluctuations
of the probability $\tau$ of no return 
into a given entrance channel for wave scattering 
from disordered systems. For zero-dimensional 
("quantum chaotic") and quasi one-dimensional
systems with broken time-reversal invariance
we derive explicit formulas for the distribution of $\tau$,
and investigate particular cases. 
Finally, relating $\tau$ to 
violation of S-matrix unitarity induced by internal dissipation,
 we use the same quantity to identify the
Anderson delocalisation transition as the phenomenon
of spontaneous breakdown of S-matrix unitarity.

\end{abstract}


Various aspects of chaotic wave scattering in the presence of
absorption or internal losses
attracted a considerable attention in recent years\cite{1}-\cite{fr}.
In the general case a convenient framework for 
extracting universal properties of the 
corresponding S-matrix is provided by the method of 
effective non-Hermitian Hamiltonian ${\cal
H}=H-i\pi \sum_{a=1}^MW_a\otimes W_a^{\dagger},$
in terms of which the energy-dependent element  
$S_{ab}$ of the $M\times M$ scattering
matrix $\hat{S}$ is expressed as:
\begin{equation}\label{1}
S_{ab}=\delta_{ab}-2i\pi W^{\dagger}_a\frac{1}{{\cal E}-{\cal H}}W_b\,,
\end{equation}
see \cite{9},\cite{10} and references therein.
Here $H$ stands for a self-adjoint Hamiltonian describing the
closed counterpart of the disordered or chaotic system under 
consideration, ${\cal E}$ stands for the energy of incoming waves
 and the (energy-independent)
vectors $W_a$, $a=1,2,...M$ contain matrix elements coupling
the internal motion to one of open $M$ channels. As is easy to verify
such a construction ensures the unitarity of the scattering matrix: 
$S^{\dagger}{S}={\bf 1}_M$, provided the energy ${\cal E}$ takes
only real values. When one allows for energy parameter to have 
nonzero imaginary part $\epsilon=\mbox{Im}{\cal E}>0$, 
the $S-$matrix unitarity is immediately
lost: $S^{\dagger}{S}<{\bf 1}_M$. 
Physically the parameter $\epsilon$ stands for a uniform damping
inside the system, and is responsible for the fact of 
losses of the outgoing flux of the particles as compared to the
incoming flux. The balance between the two fluxes is precisely the physical
mechanism standing behind the $S-$ matrix unitarity. 

In the present paper we concentrate on
the "probability of no return" (PNR), which is defined
as the quantum-mechanical probability for a
particle entering the system via a given channel $a$ never exit
through the same channel. This quantity is well-defined for a given
realization of disorder and will show sample-to-sample fluctuations
whose statistics we are going to study.  
In the case of no internal dissipation PNR is
the same as the probability to exit via any of the remaining
channels, known as the transmission probability
$\sum_{b\ne a}|S_{ab}|^2\equiv 1-|S_{aa}|^2$. 
For a system with absorption the last equality is violated, 
and we keep the notation $\tau_a=1-|S_{aa}|^2$ for 
the PNR related to the 
reflection probability $R_a$ in the same channel as $\tau_a=1-R_{a}$. 
In particular, if only a single open channel is attached to our
disordered system and the boundaries are purely reflecting,
 then neglecting dissipation trivially 
results in $\tau\equiv 0$. The nontrivial statistics of
$\tau$ then arises solely due to an absorption, and for small absorption the 
PNR value $\tau\simeq 2\epsilon \tau_{W}$\cite{2,2a,Doron}, where $\tau_W$ is the
so-called Wigner delay time intensively studied in recent
years, see \cite{2a,3},\cite{9},\cite{tsamp}-\cite{td2}
and references therein.

It is convenient to write explicitly the normalisation
of the channel vectors as
$W^{\dagger}_aW_a=\gamma_a/\pi$, and assume
that different channel vectors are orthogonal:
$W^{\dagger}_aW_b=0$ for $ a\ne b$. In fact, one should remember that the
effective strength of every open
channel is more appropriately characterised by the so-called "transmission
coefficients" \cite{10} (also known as the "sticking
probabilities"): $T_a=1-|\langle S_{aa}\rangle|^2$, related to bare
couplings $\gamma_a$ by 
\begin{equation}\label{eff}
\frac{1}{T_a}=\frac{1}{2}\left(1+g_a\right),
\quad g_a=\frac{1}{2\pi\nu}(\gamma_a+\gamma_a^{-1}),
\end{equation}
where $\nu$ stands for the mean spectral
density. Here and afterwards the angular brackets stand for the 
disorder-averaged value of the quantities. 
Two limiting cases $T_a=1$ and $T_a=0$ correspond
to situations of perfectly coupled and decoupled (closed) channel
$a$, respectively.

The starting point of our analysis is based on the following
convenient representation for the  
diagonal elements of the scattering matrix:
%\cite{note}:
\begin{equation}\label{starting}
S_{aa}=\frac{1-iG_{a}}{1+iG_{a}}\quad,\quad G_a=\pi W^{\dagger}_a
\frac{1}{{\cal E}-{\cal H}_{a}}W_a\,,
\end{equation}
where ${\cal H}_{a}=H-i\pi\sum_{b\ne a}^M W_b\otimes
W_b^{\dagger}$.
In this way we reduce our problem to investigating statistics
of the diagonal entries of the resolvent $G_a$ of the "reduced-rank"
non-Hermitian operator ${\cal H}_a$ independent of the vector $W_a$.
In particular, for the single-channel case $M=1$ the operator ${\cal H}_a$
will not contain the channel vector $W$ at all, and will be therefore
a self-adjoint one: ${\cal H}=H$. 

The statistics of the diagonal entries of the resolvent of a random
self-adjoint Hamiltonian $H$ describing the motion of a 
quantum particle in a static
disorder was discussed in much detail by Mirlin and
Fyodorov\cite{11} in the framework of the supermatrix nonlinear 
$\sigma-$ model\cite{12}. In particular, for the case of systems 
with broken time reversal invariance they were able to find a very
compact representation for the joint probability density ${\cal
P}(u,v)$ of the real $u=\mbox{Re} G_a$ and imaginary $v=\mbox{Im} G_a$ 
parts of the quantity $G_a$, assuming normalisation
$W^{\dagger}_aW_a=1/\pi$. Physically the variable $v$ is the most 
important and being the local density of states enjoyed thorough 
investigations, see\cite{mf},\cite{LDOS} and references therein.

In fact, it is quite straightforward to incorporate non-Hermitian
part of the Hamiltonian ${\cal H}_a$ into the method, as was already partly 
done in \cite{13} where statistics of $\mbox{Im}G_a$ was addressed as 
describing fluctuations of the photodissociation cross-section.

According to \cite{11} the function ${\cal P}(u,v)$ is given, for the
centre of spectrum $\mbox{Re}{\cal E}=0$, by
${\cal P}(u,v)=\frac{1}{4\pi v^2}P(x)$, where $x$ stands for the
combination $x=(u^2+v^2+1)/2v$, and the function $P(x)$ is given by 
\begin{equation}\label{mif}
P(x)=\hat{\bf L}
F(x),\quad F(x)=
\int_{-1}^{1}\frac{d\lambda}{x-\lambda}{\cal F}(x,\lambda)\,,
\end{equation}      
where we introduced the (Legendre) operator
 $\hat{\bf L}=\frac{d }{d x}(x^2-1)
\frac{d }{d x}$.
The particular form of the function ${\cal F}(x,\lambda)$
depends crucially on the effective spatial dimension of the underlying
disordered system, and is, for example, quite different for zero-dimensional
systems ("quantum chaos") and for "diffusive" extended quasi-one
dimensional, or higher dimensional systems. 
We will give explicit analysis of several
physical possibilities later on in the paper. 

Having at our disposal the expression for ${\cal P}(u,v)$ 
we can relate the PNR distribution ${\cal P}(\tau_a)$  
to the function $P(x)$. 
 After a set of algebraic transformations 
we find the following attractively simple formula:
\begin{equation}\label{simple}
{\cal P}(\tau_a)=\frac{1}{\pi
\tau_a^2}\int_{0}^{\pi}P[x(\tau_a,\theta)] d\theta
\end{equation}
\[
x(\tau_a,\theta)=\left(\frac{2}{\tau_a}-1\right)
\left(\frac{2}{T_a}-1\right)
+4\cos{\theta}
\frac{\sqrt{(1-\tau_a)(1-T_a)}}{\tau_aT_a}\,.
\]
Now we proceed with a separate analysis of a few physical
situations possible in disordered systems. In all cases
we assume time reversal symmetry to be broken.  
  
{\bf I. "Zero-dimensional" quantum chaotic system.}\\
We assume that the disordered region is coupled to $M$
scattering channels characterized by effective coupling
constants $g_1,...,g_M$, see eq.(\ref{eff}), with $g_1$
corresponding to the chosen entrance channel. The strength of uniform
damping will be characterized by the parameter
$\eta=2\pi\epsilon/\Delta$, where $\Delta$ stands for the mean level
spacing generated by the Hermitian Hamiltonian $H$.
According to the standard argumentation, $H$ can be effectively
replaced by a large $N\times N$ random Hermitian matrix taken
from the Gaussian Unitary Ensemble, see e.g.\cite{9,10}. Then in the 
limit of large enough $N\gg M$ the function
${\cal F}(x,\lambda)$ depends on the remaining $M-1$
coupling constants, as well as on the effective damping
$\eta$ as \cite{13}:
\begin{equation}\label{0dim}
{\cal F}(x,\lambda)=\prod_{a=2}^M\frac{g_a+\lambda}
{g_a+x}e^{-\eta(x-\lambda)}\,. 
\end{equation}
The function $F(x)$ in Eq.(\ref{mif})  can be found
in a closed form for any $M$ as one gets, in fact, a simple
recursion relating $F_M(x)$ to $F_{M-1}(x)$. Here we restrict ourself
mainly to the cases of one and two open channels $M=1,2$:
\begin{equation}\label{1a}
F_1(x)=\int_{x-1}^{x+1}\frac{du}{u}e^{-\eta u},\,\,
%\end{equation} 
%\begin{equation}\label{2a}
F_2(x)=F_1(x)-2\frac{e^{-\eta x}}{g_2+x}\frac{\sinh{\eta}}{\eta}\,.
\end{equation} 
The distribution ${\cal P}(\tau)$ for $M=1$ is then equal to (cf. \cite{3})
\begin{eqnarray}\label{res1}
\nonumber &&{\cal
P}(\tau)=\frac{2}{\tau^2}e^{-\eta A}\left\{I_0(\eta
B)\left[\sinh{\eta}(\eta A-1)
+\eta\cosh{\eta}\right] 
\right.\\  &-& \left. \eta B\sinh{\eta}I_1(\eta
B)\right\},\,\, A=\left(\frac{2}{\tau}-1\right)
\left(\frac{2}{T_1}-1\right),
\\ \nonumber && B=4\frac{\sqrt{(1-\tau)(1-T_1)}}{\tau T_1}\,\,,
\end{eqnarray}
where $I_0(z),I_1(z)$ stand for the modified Bessel functions
of the respective order. For particular case of perfectly coupled channel
$T_1=1$ Eq.(\ref{res1}) reduces to the 
formula 
\begin{equation}\label{perf1}
{\cal P}_1(\tau)=\frac{1}{\tau^3}e^{-2\eta/\tau}
\left[\tau\left(1+2\eta-e^{2\eta}\right)+2\eta\left(e^{2\eta}-1\right)
\right]\,
\end{equation}
derived earlier \cite{2} with a very different method.

The function ${\cal P}(\tau)$ for $M=2$ can 
be obtained straightforwardly, but the
general formula is too long, and we restrict our discussion by a few
particular cases. First of all, when dissipation is absent
($\eta=0$) we recover the exact distribution of the transmission
probability found earlier in \cite{14,15} by rather different methods.
Next case to be considered is that of a lossy system coupled to 
two perfectly open channels $g_1=g_2=1$:
\begin{equation}\label{perf2}
{\cal P}_2(\tau)={\cal P}_1(\tau)+\frac{1-e^{-2\eta/\tau}}{\eta}
\left[\frac{1}{2}+\frac{\eta}{\tau}+\frac{2\eta^2}{\tau^2}-
\frac{2\eta^2}{\tau^3}\right]
\end{equation}
where ${\cal P}_1(\tau)$ is given in Eq.(\ref{perf1}).
In fact, it is not difficult to find a similar recursive formula 
relating ${\cal P}(\tau)$ for $M$ perfectly coupled channels to
the same function for $M-1$ perfect channels. We do not give that
formula, apart from the simplest case of no dissipation:
\[
{\cal P}_M(\tau)={\cal
P}_{M-1}(\tau)+\left[(M-1)\tau^{M-2}-(M-2)\tau^{M-3} \right]\,,
\]
which immediately yields ${\cal P}_M(\tau)=(M-1)\tau^{M-2},\,M\ge 1$.
This formula (as well as its counterpart 
${\cal P}(\tau)\propto \tau^{\frac{M-3}{2}}$ for {\it preserved} time
reversal invariance), in fact, follows from the known 
distribution of $1\times 1$ subunitary block of 
random unitary scattering matrices, see \cite{my}.

{\bf II. Quasi-1D systems.} Consider a single channel attached to 
one edge of a piece
of quasi one-dimensional disordered metal of length $L$, 
with the opposite edge being in
contact with perfectly conducting lead of very many channels. 
When the internal dissipation is absent, the function
$F(x)$ was found by Mirlin \cite{mirlin}:
\begin{eqnarray}\label{mirlin}
&& F(x)=\ln{\frac{x+1}{x-1}}\\ \nonumber
&-&2\int_0^{\infty}\frac{dk k}{1+k^2}\tanh{\left(\frac{\pi k}{2}\right)}
P_{-\frac{1}{2}+\frac{ik}{2}}(x)e^{-t(k^2+1)/4}\,,
\end{eqnarray}
where the dimensionless parameter $t=L/\xi$ 
is the sample length $L$ measured in units of the
localisation lengths $\xi$. 
The (real) functions $P_{\nu}(x),\,\nu=-1/2+ik/2$ are known as conical
functions, and represent a special 
case of Legendre functions. As such they satisfy: $\hat{\bf
L}P_{\nu}(x)=\nu(\nu+1)P_{\nu}(x)$. This observation immediately
yields the following expression for the PNR distribution:
\begin{eqnarray}\label{1d1}
&&{\cal P}(\tau)=\frac{1}{2\tau^2}I\left(t;\frac{2}{\tau}-1\right),
\\ &&I(t;x)=\int_0^{\infty}{dk k}\tanh{\left(\frac{\pi k}{2}\right)}
P_{-\frac{1}{2}+\frac{ik}{2}}(x)e^{-t(k^2+1)/4},
\end{eqnarray}
where we assumed, for simplicity, that the selected single channel is
perfectly coupled to the scattering medium. Surprisingly, 
this distribution is practically the same as the distribution
of the reflection coefficient from a piece of strictly one-dimensional
medium obtained long ago in the framework of the Berezinskii 
technique \cite{Abr}. In particular, for any value of the parameter 
$t$ the distribution displays a log-normal far tail
corresponding to very small PNR values $\tau\to 0$.
To find it for $t\ll 1$ one needs an asymptotic of the conical
functions for large arguments $x\gg 1$ which we borrow from
Eq.(50) of the paper\cite{mirlin}. A calculation very similar to that
presented in \cite{mirlin} yields:
\begin{equation}\label{logtail}
{\cal P}(\tau \ll 1)\simeq
\frac{1}{2\sqrt{2}t}\frac{\sqrt{-\ln{\tau}}}{\tau} 
\exp\left\{-\frac{1}{4t}\ln^2{\tau}\right\},\quad t\ll |\ln{\tau}|
\end{equation}
This log-normal tail is related to the presence of the anomalously 
localised states\cite{rep}.
In the opposite case of very long samples $t\gg 1$ the PNR values
are exponentially small due to localisation phenomenon and the
distribution is purely log-normal:
\begin{equation}\label{logtail1}
{\cal P}(\tau)\simeq
\frac{1}{2\tau}\frac{1}{\sqrt{\pi t}} 
\exp\left\{-\frac{1}{4t}\left(t+\ln{\tau}\right)^2\right\},
\quad t\sim |\ln{\tau}|\gg 1
\end{equation}

Let us turn our attention now to the case of a quasi-1D 
disordered sample with a nonvanishing internal dissipation $\epsilon>0$,
assuming second edge of the sample to be impenetrable for waves. 
The scaling physical
parameter controlling the role of dissipation is then given by\cite{mf}
$\delta=\pi\rho\epsilon \xi$, with $\rho$
standing for the mean spectral density.
 This is just the dissipation $\epsilon$ measured in units of
the mean level spacing for a sample whose length is $\xi$.
The most interesting regime is that of small $\delta\ll 1$.
In that limit the function ${\cal F}(x,\lambda)$ turned out to be
independent of $\lambda$, whereas the $x-$ dependence persists in a
form of the scaling combination $y=2\delta x$, i.e. ${\cal F}(x,\lambda)
\to \tilde{F}(2\delta x)$. This implies that the relevant values 
of parameters are $ 2/\tau\simeq x \sim \delta^{-1}\gg 1$. 
The latter condition immediately results
in the formula $F(x)\to 4\delta \tilde{F}(y)/y$, and also converts
the Legendre operator $\hat{\bf L}$ to $\hat{\bf L}_y=\frac{d}{dy}
y^2 \frac{d}{dy}$. Let us note that 
the emerging PNR distribution yields, in fact, the distribution of  
the Wigner delay time via the relation 
$\tau_W=2\pi\rho \xi/y$. 

The expression for the function $\tilde{F}(y)$ is known
explicitly\cite{fm}:
$\tilde{F}(y)=\tilde{F}_{\infty}(y)+\tilde{F}_t(y)$, where  
\begin{eqnarray}\label{fm}
\tilde{F}_t(y)=
\frac{4}{\pi}{y}^{\frac{1}{2}}
\int_0^{\infty}\frac{dk k}{1+k^2}\sinh{\frac{\pi k}{2}}
K_{ik}\left(2{y}^{\frac{1}{2}}\right)e^{-t(k^2+1)/4}
\end{eqnarray}
and 
$\tilde{F}_{\infty}(y)=2{y}^{\frac{1}{2}}K_{1}\left(2{y}^{\frac{1}{2}} 
\right)$, with $K_{\nu}(u)$ standing for the Macdonald function.

For the case of very short $(t\ll 1)$ sample the function 
$\tilde{F}(y)$ is known to be approximated
by $\exp{-(ty)}$\cite{fm}.
A simple calculation then yields the distribution
${\cal P}(\tau) =\left[(4t\delta)^2/\tau^3\right]
\exp\left\{ -4t\delta/\tau\right\}$. Realising that $2t\delta\equiv
\eta$ we see that the distribution coincides with the weak
absorption limit of Eq.(\ref{perf1}). As expected, the same
distribution follows from that of the Wigner delay time\cite{9}. 

In the opposite limit of very long samples $t\to\infty$ only first term
survives, and by noticing that $\hat{\bf L}_y\left[
\tilde{F}_{\infty}(y)/y\right]=\tilde{F}_{\infty}(y)$, we find the
corresponding PNR distribution:
\begin{equation}\label{1d3}
{\cal
P}(\tau)=\frac{16}{\delta}\left(\frac{\delta}{\tau}\right)^{5/2} 
K_1\left(4\sqrt{\delta/\tau}\right).
\end{equation}
Although the typical value of $\tau$ is of the order of $\delta$,
the moments $\langle \tau^m\rangle$ do not exist for $m\ge 1$
because of the powerlaw tail ${\cal P}(\tau\gg \delta)\propto \tau^{-2}$. 
A similar tail was found in the distribution of 
the total reflection coefficient
from multichannel long disordered 1D sample in \cite{7}, and
is also typical for the Wigner delay time distribution in purely 1D
system\cite{td2}. 
Negative moments of $\tau $ are equal to
$\langle \tau^{-k}\rangle=(4\delta)^{-k}k!(k+1)!$. Note that  
they differ from the corresponding moments in purely 1D 
system\cite{td2} by extra factorial factor $(k+1)!$, reminiscent
of similar relations between other quantities in $1D$ and quasi $1D$
\cite{mf}.

Finally, in the case of strong absorption $\delta\gg 1$ in a long wire $t\to
\infty$ the function ${\cal
F}(x,\lambda)=\exp\{-\sqrt{\delta}(x-\lambda)\}$\cite{12} and the  
resulting distribution ${\cal P}(\tau)$ coincides with that
given by Eq.(\ref{perf1}), with $\eta$ replaced by $\sqrt{\delta}$.

{\bf III. Behaviour at the Anderson transition}. 

Lets us shortly discuss a possible qualitative behaviour of 
the PNR $\tau$ in a scattering system formed by a single perfect channel
attached to a $d-$ dimensional disordered
sample at the vicinity of the point of the Anderson
delocalisation transition $\alpha_c$ (the mobility edge). Here we denote by $\alpha$ an 
effective parameter which controls the transition in the infinite
sample, with states
being localised (extended) for $\alpha >\alpha_c$ 
(respectively, $\alpha <\alpha_c$). 

Our arguments are based on a picture
of the transition as described in terms of a functional order parameter
developed in detail in \cite{11}, see also earlier results in\cite{12}
and \cite{trans}. For a sample of finite size $L$ the PNR is
 a function of three parameters: $\epsilon,L,\alpha$.
According to the suggested scenario, 
the behaviour of the function ${\cal F}(x,\lambda)$ in the insulating
phase in the weak absorption
limit $\delta\propto \epsilon \xi^d\to 0$ 
is expected to be reminiscent of that described above for 
the one-dimensional case, i.e. ${\cal F}(x,\lambda)
\to \tilde{F}(2\delta x)$, and the function $\tilde{F}(y)$
decays to zero for $y\gg 1$. Then it is natural to expect that
 all negative PNR moments in the infinite volume limit 
$L\to \infty$ are to scale as 
$\langle \tau^{-k} \rangle \sim \epsilon^{-k} \xi^{-d k}$, where
 $\xi$ is the localisation length diverging in the vicinity of the 
mobility edge. 

In contrast, in the delocalized phase the function  
${\cal F}(x,\lambda)$ is expected to remain a non-trivial function of
both $x$ and $\lambda$ even when $\epsilon\to 0$, provided
the latter limit is taken {\it after} the infinite volume limit $L\to \infty$. 
This should immediately result in a finite-width  
distribution ${\cal P}(\tau)$ of the PNR. 
From this point of view the Anderson transition 
acquires a natural interpretation
as the phenomenon of {\it spontaneous} breakdown of S-matrix
unitarity. As long as
$\alpha\to\alpha_c$, the widths of the distribution and
properly defined (negative) PNR moments should vanish, with
some set of critical exponents. 

If, however, we take limit $\epsilon\to 0$ first, then 
for $\alpha <\alpha_c$ PNR in large but 
finite sample should scale with the system size $L$ as $\langle \tau^{-k} 
\rangle \sim C(\alpha) \epsilon^{-k} L^{-d k}$, where 
$C(\alpha)$ is expected to diverge when $\alpha \to \alpha_c$.
In some sense the behaviour of the negative moments of
the Wigner delay time defined as $\tau_W=
\lim_{\epsilon\to 0}\tau(\epsilon,L,\alpha)/2\epsilon$ 
is reminiscent of that for the inverse participation ratio\cite{rep}.
This analogy suggests a possibility for anomalous scaling $\langle \tau_W^{-k} 
\rangle \sim L^{-d r_k}$ with $r_k\ne k$ at the mobility edge
$\alpha=\alpha_c$, which would then reflect the underlying
multifractality of the wavefunctions.

It will be very interesting to perform a detailed numerical 
analysis of PNR and Wigner delay times for realistic and
well-controlled models of scattering from 
disordered systems, e.g. quantum graphs\cite{ks}
or models used in \cite{tsamp} and to
verify the suggested picture qualitatively and quantitatively
in various regimes. The statistics of PNR should be also of
experimental accessibility in microwave resonators type of experiments, 
see e.g \cite{5,Doron} and references therein.

The author is obliged to Y.Imry and U.Kuhl for useful 
discussions and to T. Kottos, A.Mirlin and D.Savin for valuable comments.  
Financial support by Brunel University VC grant is gratefully acknowledged.


\begin{references}
\bibitem{1} E.Kogan et al.
 Phys.Rev. E {\bf 61}, R17 (2000)
\bibitem{2} C.W.J. Beenakker, P.W.Brouwer, Physica E {\bf 9},
463 (2001)  
\bibitem{2a} S.A. Ramakrishna, N. Kumar, Phys. Rev.B {\bf 61},
3163 (2000)
\bibitem{3} 
D.V.Savin, H.-J.Sommers cond-mat/0303083; 
\bibitem{4}
I. Rozhkov, Y.V. Fyodorov, and R.L. Weaver, 
Phys. Rev. E {\bf 68}, 016204 (2003). 
\bibitem{5} 
R.Sch\"{a}fer et al., J. Phys. A: Math. Gen. {\bf 36}, 3289 (2003);   
R.A. Mendez-Sanchez et al. cond-mat/0305090 
\bibitem{Doron} E.Doron et al. Phys. Rev. Lett. {\bf 65}, 3072 (1990)
\bibitem{6}O. Entin-Wohlman et al. Phys. Rev. Lett. {\bf 88}, 166801
\bibitem{7}T. Sh. Misirpashaev, C. W. J. 
Beenakker, JETP Lett. {\bf 64} , 319 (1996);
N.A. Bruce, J.T.Chalker, J.Phys.A:Math.Gen. {\bf 29},
3761 (1996)
\bibitem{fr} V.Freilikher et al.
Phys.Rev. B {\bf 50}, 6017 (1994); 
P. Pradhan, N. Kumar , ibid. {\bf 50}, 9644 (1994);
M.Titov and C.W.J. Beenakker, Phys. Rev. Lett. {\bf 85}, 3388 (2000)
\bibitem{9}Y.V. Fyodorov, H.-J. Sommers, J. Math. Phys. {\bf 38} (4), 1918
(1997).
\bibitem{10}J.J.M.Verbaarschot et al. Phys. Rep. {\bf 129}, 367 (1985). 
\bibitem{tsamp} T.Kottos, M.Weiss, Phys. Rev. Lett. {\bf 89}, 056401
(2002); A. Ossipov et al. , Phys.Rev. B {\bf 61}, 11411 (2000) and
 Europhys. Lett. {\bf 62}, 719 (2003).
\bibitem{td1} P.W.Brouwer et al. Phys.Rev.Lett. {\bf 78}, 4737 (1997)
\bibitem{td2} C.Texier, A.Comtet, Phys.Rev.Lett. {\bf 82}, 4220
(1999); S. K. Joshi et al., Phys.Rev.B {\bf 58}, 1092 (1998);
J.Heinrichs, Phys.Rev.B {\bf 65}, 075112 (2002)
\bibitem{11} A.D. Mirlin, Y.V.Fyodorov, Phys.Rev.Lett. {\bf 72},
526 (1994); J. Phys. I France, 
{\bf 4}, 655 (1994) 
\bibitem{12}K.B. Efetov. {\it Supersymmetry in disorder and 
chaos} (Cambridge University Press, 1997).
\bibitem{mf} A.D.Mirlin, Y.V.Fyodorov, Europhys. Lett. {\bf 25}, 669
(1994); C.W.J. Beenakker, Phys.Rev.B {\bf 50}, 15170 (1994)  
\bibitem{LDOS} H.Schomerus et al., Phys.Rev.B {\bf 65}, 121101
(2002); M.Titov, H.Schomerus , ibid. {\bf 67}, 024410
\bibitem{13}Y. V. Fyodorov, Y.  Alhassid, Phys. Rev. A, 
{\bf 58}, R3375(1998).  
\bibitem{14} V.N.Prigodin et al.
Phys.Rev.B {\bf 51}, 17223 (1995) 
\bibitem{15} P.W. Brouwer, C.W.J. Beenakker, 
Phys.Rev.B {\bf 50}, 11263 (1994) 
\bibitem{my}Y.V.Fyodorov, H.-J.Sommers, J.Phys.A:Math.Gen. 
{\bf 36},3303 (2003); see Eq.(69) of the paper;
 W.A.Friedman, P.A.Mello, J.Phys.A:Math.Gen. 
{\bf 18},425 (1985) 
\bibitem{mirlin} A.D.Mirlin,  Phys.Rev.B {\bf 53}, 1186 (1996)
\bibitem{rep} A.D.Mirlin, Phys.Rep. {\bf 326}, 259 (2000)
\bibitem{Abr} A.A.Abrikosov, Solid State Comm. {\bf 37}, 997 (1981)
\bibitem{fm} Y.V.Fyodorov, A.D.Mirlin, Int.J.Mod.Phys.B 
{\bf 8}, 3795 (1994)
\bibitem{trans} M.R.Zirnbauer Phys.Rev.B {\bf 34},6394 (1986)
A.D.Mirlin, Y.V. Fyodorov, Nucl.Phys. B [FS] {\bf 366}, 507  (1991) 
\bibitem{ks}T. Kottos, U. Smilansky, 
Phys.Rev.Lett. {\bf 85}, 968 (2000);
J.Phys.A:Math.Gen {\bf 36}, 3501 (2003)


\end{references}
\end{document}